\documentclass[prb,aps,superscriptaddress,reprint,longbibliography]{revtex4-2}
\usepackage{siunitx}
\usepackage{graphicx}
\usepackage{amsmath}
\usepackage[english]{babel}
\usepackage[colorlinks=true,urlcolor=blue,linkcolor=blue,citecolor=blue]{hyperref}
\usepackage{mathptmx}
\usepackage{gensymb}

\begin{document}

\title{Benchmarking the integration of hexagonal boron nitride crystals and thin films into graphene-based van der Waals heterostructures}

\author{Taoufiq Ouaj} 
\affiliation{2nd Institute of Physics and JARA-FIT, RWTH Aachen University, 52074 Aachen, Germany}

\author{Christophe Arnold} 
\affiliation{GEMaC, UVSQ, CNRS, Université Paris Saclay, France}

\author{Jon Azpeitia} 
\affiliation{Instituto de Ciencia de Materiales de Madrid (ICMM-CSIC), Sor Juana Inés de la Cruz 3, 28049 Madrid, Spain}

\author{Sunaja Baltic}
\affiliation{2nd Institute of Physics and JARA-FIT, RWTH Aachen University, 52074 Aachen, Germany}

\author{Julien Barjon} 
\affiliation{GEMaC, UVSQ, CNRS, Université Paris Saclay, France}

\author{Jose Cascales}
\affiliation{Instituto de Ciencia de Materiales de Madrid (ICMM-CSIC), Sor Juana Inés de la Cruz 3, 28049 Madrid, Spain}

\author{Huanyao Cun}
\affiliation{Physik-Institut, University of Zürich, Zürich, Switzerland}

\author{David Esteban}
\affiliation{Instituto de Ciencia de Materiales de Madrid (ICMM-CSIC), Sor Juana Inés de la Cruz 3, 28049 Madrid, Spain}

\author{Mar Garcia-Hernandez}
\affiliation{Instituto de Ciencia de Materiales de Madrid (ICMM-CSIC), Sor Juana Inés de la Cruz 3, 28049 Madrid, Spain}

\author{Vincent Garnier}
\affiliation{INSA Lyon, Universite Claude Bernard Lyon 1, CNRS, MATEIS, UMR5510,69621 Villeurbanne, France}

\author{Subodh K. Gautam} 
\affiliation{GEMaC, UVSQ, CNRS, Université Paris Saclay, France}

\author{Thomas Greber}
\affiliation{Physik-Institut, University of Zürich, Zürich, Switzerland}

\author{Said Said Hassani} 
\affiliation{GEMaC, UVSQ, CNRS, Université Paris Saclay, France}

\author{Adrian Hemmi}
\affiliation{Physik-Institut, University of Zürich, Zürich, Switzerland}

\author{Ignacio Jimenéz}
\affiliation{Instituto de Ciencia de Materiales de Madrid (ICMM-CSIC), Sor Juana Inés de la Cruz 3, 28049 Madrid, Spain}

\author{Catherine Journet}
\affiliation{Universite Claude Bernard Lyon 1, CNRS, LMI UMR 5615, Villeurbanne, F-69100, France}

\author{Paul Kögerler} 
\affiliation{Institute of Inorganic Chemistry, RWTH Aachen University, 52074 Aachen, Germany}
\affiliation{Peter Grünberg Institute (PGI-6), Forschungszentrum Jülich, 52425 Jülich, Germany}

\author{Annick Loiseau}
\affiliation{Université Paris Saclay, ONERA, CNRS, Laboratoire d’Etude des Microstructures, 92322 Chatillon France}

\author{Camille Maestre}
\affiliation{Universite Claude Bernard Lyon 1, CNRS, LMI UMR 5615, Villeurbanne, F-69100, France}

\author{Marvin Metzelaars}
\affiliation{2nd Institute of Physics and JARA-FIT, RWTH Aachen University, 52074 Aachen, Germany}
\affiliation{Institute of Inorganic Chemistry, RWTH Aachen University, 52074 Aachen, Germany}

\author{Philipp Schmidt}
\affiliation{2nd Institute of Physics and JARA-FIT, RWTH Aachen University, 52074 Aachen, Germany}

\author{Christoph Stampfer}
\affiliation{2nd Institute of Physics and JARA-FIT, RWTH Aachen University, 52074 Aachen, Germany}
\affiliation{Peter Gr\"unberg Institute (PGI-9) Forschungszentrum J\"ulich, 52425 J\"ulich, Germany}

\author{Ingrid Stenger} 
\affiliation{GEMaC, UVSQ, CNRS, Université Paris Saclay, France}

\author{Philippe Steyer} 
\affiliation{INSA Lyon, Universite Claude Bernard Lyon 1, CNRS, MATEIS, UMR5510,69621 Villeurbanne, France}

\author{Takashi Taniguchi}
\affiliation{Research Center for Materials Nanoarchitectonics, National Institute for Materials Science,  1-1 Namiki, Tsukuba 305-0044, Japan}

\author{Bérangère Toury}
\affiliation{Universite Claude Bernard Lyon 1, CNRS, LMI UMR 5615, Villeurbanne, F-69100, France}

\author{Kenji Watanabe}
\affiliation{Research Center for Electronic and Optical Materials, National Institute for Materials Science, 1-1 Namiki, Tsukuba 305-0044, Japan}

\author{Bernd Beschoten} 
\email{bernd.beschoten@physik.rwth-aachen.de}
\affiliation{2nd Institute of Physics and JARA-FIT, RWTH Aachen University, 52074 Aachen, Germany}

\begin{abstract}
  We present a benchmarking protocol that combines the characterization of boron nitride (BN) crystals and films with the evaluation of the electronic properties of graphene on these substrates.
  Our study includes hBN crystals grown under different conditions (atmospheric pressure high temperature, high pressure high temperature, pressure controlled furnace) and scalable BN films deposited by either chemical or physical vapor deposition (CVD or PVD). 
  We explore the complete process from boron nitride growth, over its optical characterization by time-resolved cathodoluminescence (TRCL), to the optical and electronic characterization of graphene by Raman spectroscopy after encapsulation and Hall bar processing.
  Within our benchmarking protocol we achieve a homogeneous electronic performance within each Hall bar device through a fast and reproducible processing routine.
  We find that a free exciton lifetime of $\mathrm{1 \, ns}$ measured on as-grown hBN crystals by TRCL is sufficient to achieve high graphene room temperature charge carrier mobilities of $\mathrm{80,000 \, cm^2/(Vs)}$ at a carrier density of $\mathrm{|n| = \num{1e12} \, cm^{-2}}$, while respective exciton lifetimes around $\mathrm{100 \, ps}$ yield mobilities up to $\mathrm{30,000 \, cm^2/(Vs)}$.
  For scalable PVD-grown BN films, we measure carrier mobilities exceeding $\mathrm{10,000 \, cm^2/(Vs)}$ which correlates with a graphene Raman 2D peak linewidth of $\mathrm{22 \, cm^{-1}}$.  
  Our work highlights the importance of the Raman 2D linewidth of graphene as a critical metric that effectively assesses the interface quality (i.e. surface roughness) to the BN substrate, which directly affects the charge carrier mobility of graphene.
  Graphene 2D linewidth analysis is suitable for all BN substrates and is particularly advantageous when TRCL or BN Raman spectroscopy cannot be applied to specific BN materials such as amorphous or thin films. This underlines the superior role of spatially-resolved spectroscopy in the evaluation of BN crystals and films for the use of high-mobility graphene devices.

\end{abstract}

\maketitle

\section{Introduction}
\begin{figure*}[t]
  \begin{center}
  \includegraphics{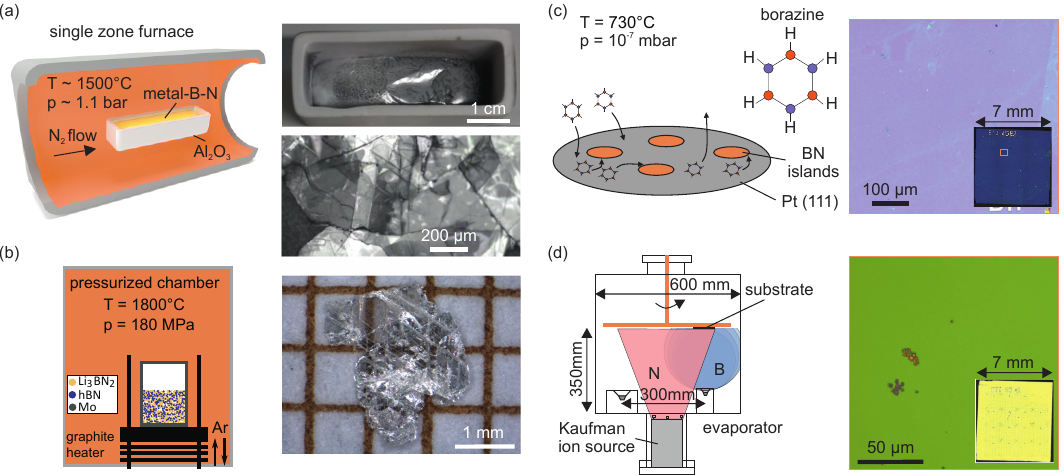}
  \end{center}
  \caption{Overview of growth techniques for hBN crystals and films. 
  {\bf(a)} Atmospheric pressure high temperature growth process with schematic of a gas flow furnace. Optical images of the resulting hBN crystals on top of the iron ingot and of the crystals after detaching from the iron ingot. 
  {\bf (b)} Schematic of the growth of hBN crystals in a pressure controlled furnace and optical image of hBN crystal after dissolution of $\mathrm{Li_3BN_2}$.
  {\bf(c)} CVD growth of hBN on Pt(111) substrate and an optical image of the transferred hBN film on a $\mathrm{Si/SiO_2}$ wafer piece.
  {\bf (d)} Schematic of the physical vapor deposition growth setup for BN and optical image of a $\mathrm{Si/SiO_2}$ wafer with the PVD-grown film.}
  \label{fig: growth}
\end{figure*}

Boron nitride (BN) with its remarkable thermal stability, chemical inertness and robust mechanical properties has long been used for various applications~\cite{Meng2019Feb, Schue2016Dec, Backes2020Jan, Naclerio2023Feb}.
It has been demonstrated that hexagonal boron nitride (hBN) is of particular importance for applications in 2D material systems, exhibiting properties crucial for photonics and optoelectronics, such as efficient deep UV emissions~\cite{Watanabe2004Jun, Kubota2007Aug} and quantum photonics capabilities~\cite{Bourrellier2016Jul, Grosso2017Sep, Martinez2016Sep, Fournier2021Jun}.
The high thermal conductivity~\cite{Lindsay2011Oct,Yuan2019May}, the large electronic bandgap~\cite{Watanabe2004Jun}, and the ultra-flat and inert surface \cite{Xue2011Apr} are important prerequisites for the use as a substrate for other 2D materials or for interface engineering~\cite{Woods2014Jun,Britnell2012Mar,Sharpe2019Aug,Tebbe2023Apr,Tebbe2024May}.
2D materials encapsulated in hBN allow for record-breaking charge carrier mobilites in graphene~\cite{Dean2010Oct, Wang2013Nov,Banszerus2015Jul,Onodera2020Jan, Sonntag2020Jun,Ouaj2023Sep}, high electronic and optical quality in transition metal dichalcogenides (TMDs)~\cite{Ajayi2017Jul,Ye2018Sep, Raja2019Sep, Cadiz2017May, Ersfeld2020May,Shi2020Jul} or, for example, bilayer graphene quantum devices with ultra-clean tunable bandgaps~\cite{Eich2018Aug,Banszerus2018Aug,Icking2022Nov,Icking2024Sep}.

In fundamental research, hBN flakes exfoliated from bulk crystals grown either at high temperature and high pressure (HPHT)~\cite{Taniguchi2007May,Zhigadlo2014Sep,Fukunaga2022Nov} or at atmospheric pressure and high temperature (APHT)~\cite{Kubota2007Aug, Kubota2008Mar, Hoffman2014May, Edgar2014Oct, Liu2017Sep, Liu2018Sep, Zhang2019Nov, Li2020Jun, Li2021Apr, Li2020, Li2021, Zhang2021May, Ouaj2023Sep} are employed for high-quality device fabrication due to their superior crystal quality. 
The synthesis of high-quality hBN crystals in a pressure-controlled furnace (PCF) is a recent development that offers new opportunities for improving  material quality~\cite{Li2020Feb,Maestre2022May}.
hBN single crystals are small, a few millimeters at most, and therefore do not meet industrial manufacturing requirements. The transition of BN from the use in fundamental research to industrial applications requires process development capable of providing large area single crystal or polycrystalline films that meet both device requirements and high volume production needs. 

Techniques like chemical vapor deposition (CVD) \cite{Chubarov,Jang2016May,Lee2018Nov,Asgari2022Jul,Calandrini2023Nov,Tailpied2024Aug}, metal-organic CVD (MOCVD) \cite{Kobayashi2008Nov,Kobayashi2012Apr,Li2016Jun,Jeong2019Apr}, molecular beam epitaxy (MBE) \cite{Cho2016Sep,Elias2019Jun,Heilmann2020Feb,Rousseau2024Mar}, 
and physical vapor deposition (PVD) \cite{Caretti2011Aug,Jimenez2012Mar} are under development, offering potential platforms for BN substrates with sufficient interface and/or bulk qualities for the desired technological applications.
Recently, amorphous (or nanocrystalline) boron nitride (aBN), has gained interest due to its ability to be grown at room temperature on arbitrary substrates \cite{Torres2014Aug} and its low dielectric constant \cite{Hong2020Jun, Sattari-Esfahlan2023Feb, Martini2023Aug}.
Especially the full encapsulation of CVD-grown graphene in direct-grown aBN was recently reported to have promising electronic properties, showing its potential as a scalable substrate for graphene and other 2D materials~\cite{Sattari-Esfahlan2023Feb}.

The evolving diversity of available BN substrates -- from high-quality hBN crystals to hBN/aBN films -- underlines the need for comparable and meaningful characterization methods of both the crystal quality itself and the ability to be used as substrate in van der Waals heterostructures.
To assess the crystal quality, BN is mostly investigated by cathodoluminescence (CL)~\cite{Schue2016Mar, Schue2019Feb, Roux2021Oct}, photoluminescence (PL)~\cite{Cassabois2016Apr, Elias2019Jun, Rousseau2021Dec} or Raman spectroscopy \cite{Reich2005May,Schue2016Dec,Stenger2017Jun}.
Raman spectroscopy gives a rapid and non-invasive way to extract the quality of BN films and therefore is an indispensable tool to efficiently monitor the parameter tuning during optimization of growth processes.
On the other side, CL measurements and especially time-resolved cathodoluminescence (TRCL) measurements yield a much more sensitive way to evaluate the crystal quality and to gain a deeper understanding of the type of crystal defects \cite{Schue2016Mar, Schue2019Feb, Roux2021Oct}.
Here, the free exciton lifetime, which is limited by exciton-defect scattering, yields a sensitive benchmark for the bulk crystal quality.
However, CL measurements are not suitable for most scalable BN growth approaches as they are only applicable to crystalline and thick ($\mathrm{>10\, \mu m}$) hBN.
While these evaluations are highly important for benchmarking the quality of hBN crystals for optical applications with hBN as the active layer, methods to evaluate the surface quality become equally important when used as a substrate~\cite{Banszerus2017Feb}.
For example, correlations between the amount or type of defects and the surface roughness seem possible but remain a topic under investigation~\cite{Caretti2012Sep}.

Graphene, due to its exceptional high charge carrier mobility, is one of the most interesting 2D materials to be used in combination with BN.
Additionally to the interest due to its electronic properties, graphene is highly sensitive to charge disorder and surface roughness of the substrate, drastically limiting the device performance \cite{Wang2013Nov,Couto2014Oct}.
Due to both, its huge potential for future high-mobility applications and its high sensitivity to the underlying substrate, the evaluation of graphene itself on the substrate of interest is an appealing way to investigate the suitability of various BN films or crystals as a substrate for 2D materials.
Spatially-resolved confocal Raman microscopy on graphene encapsulated in BN provides a very powerful and sensitive way to directly assess strain, doping, and nm-strain variations~\cite{Lee2012Aug,Neumann2015Sep,Banszerus2017Feb,Vincent2018Dec} and directly link it to the electronic properties extracted from charge transport measurements on Hall bar devices \cite{Couto2014Oct, Onodera2020Jan, Sonntag2020Jun, Ouaj2023Sep}.

Here, we present a comprehensive evaluation of various BN substrates and present a benchmarking protocol covering the characterization of the BN as well as the evaluation of the electronic properties of exfoliated graphene on these BN substrates.
Our study includes the growth of both high-quality hBN crystals grown via APHT or in a PCF and the growth of scalable BN films via PVD or CVD (section II).
We extract the free exciton lifetime from TRCL measurements to compare the crystal quality of BN crystals and evaluate both exfoliated flakes and films via Raman spectroscopy (section~III).
Using exfoliated graphene, we fabricated dry-transferred devices on the BN substrates to assess the interface quality via spatially-resolved Raman spectroscopy (section IV).
To establish reliable benchmarks we focus on the full width at half maximum (FWHM) of the graphene Raman 2D peak, which we identify as the most sensitive benchmark for an early-stage evaluation of the suitability for a diverse set of BN substrates (section V).
Following a newly developed fabrication scheme, with a focus on rapid processing, we build Hall bar structures (section VI) to extract the charge carrier mobility at different charge carrier densities (section VII).
We demonstrate that graphene encapsulated in APHT hBN crystals compares in electronic quality to graphene encapsulated in HPHT-grown hBN crystals, reaching room temperature charge carrier mobilities around $\mathrm{80,000 \, cm^2/(Vs)}$ at a charge carrier density of $n=\mathrm{\num{1e12} \, cm^{-2}}$.
Importantly, we identify a free exciton lifetime of above $\mathrm{1 \, ns}$ to be sufficient to achieve these high charge carrier mobilities and of $\mathrm{100 \, ps}$ for charge carrier mobilities up to $\mathrm{30,000 \, cm^2/(Vs)}$.
Specifically, we demonstrate that graphene on PVD-grown nanocrystalline boron nitride with a graphene 2D peak FWHM below $\mathrm{22 \, cm^{-1}}$ consistently yields charge carrier mobilities exceeding $\mathrm{10,000 \, cm^2/(Vs)}$. 
This underscores the potential of PVD-grown BN films as scalable substrates for high-mobility graphene devices.

\section{Growth and Preparation of Boron Nitride}

In Fig.~\ref{fig: growth} the growth and preparation conditions of hBN crystals (APHT and PCF growth) and of BN thin films (PVD and CVD growth) are summarized. 

\subsection*{Atmospheric Pressure and High Temperature (APHT)}
The hBN crystals in this study were grown from an iron flux (at RWTH) or chromium-nickel (at RWTH and GEMaC) flux via the APHT method
(see Ref.~\cite{Ouaj2023Sep} for details on the growth at RWTH).
A schematic illustration of the growth setup is shown in Fig.~\ref{fig: growth}(a).
The boron source is either boron powder (RWTH) mixed with the metal pieces or a pyrolytic BN crucible (GEMaC).
The system is first  annealed at high temperature under a continuous gas flow of either $\mathrm{H_2}$ and $\mathrm{Ar}$ (RWTH) or $\mathrm{N_2}$ (GEMaC) to minimize contaminations with oxygen and carbon. 
The hBN crystal growth is started upon introduction of $\mathrm{N_2}$ while maintaining a constant pressure. 
After a soaking phase at high temperature to saturate the metal flux with B and N, the furnace is cooled down to a lower temperature at a slow rate (typically between $\mathrm{0.5^{\circ}C/h}$ and $\mathrm{4^{\circ}C/h}$). 
The system is then quickly quenched down to room temperature. 
The resulting thick hBN crystal layer is firmly attached to the underlying metal ingot, as seen in Fig.~\ref{fig: growth}(a) (right upper panel). 
The hBN crystal sheet can be detached from the metal ingot by immersion in hydrochloric acid at room temperature, see the detached crystal sheet in the lower right panel of Fig.~\ref{fig: growth}(a).
This step does not affect the quality of the hBN crystals and simplifies the further processing of the hBN crystals for exfoliation and subsequent dry-transfer \cite{Ouaj2023Sep}.
 
\subsection*{Growth in a pressure-controlled furnace (PCF)}
In the PCF method, hBN crystals are grown from the liquid phase of $\mathrm{Li_3BN_2-BN}$ at high temperature \cite{Solozhenko1997Aug, Maestre2021Oct,Maestre2022May}. 
The $\mathrm{Li_3BN_2}$ powder is pre-synthesized from $\mathrm{Li_3N}$ (Sigma Aldrich, purity  $\mathrm{> 99.5\%}$) \cite{Sahni2018May} and mixed with commercial hBN powder (20 wt\% hBN and 80 wt\% $\mathrm{Li_3BN_2}$) in a molybdenum crucible. 
Since $\mathrm{Li_3BN_2}$ is very sensitive to air and moisture, the growth preparation is performed under inert conditions and careful handling is necessary throughout the whole process. 
Both, hBN powder and crucibles are pre-treated at $\mathrm{1200^{\circ}C}$ under vacuum and an $\mathrm{Ar/H_2}$ gas mix to remove potential contaminations. 
The growth is performed in a pressure-controlled furnace (PCF)~\cite{Li2020Feb, Maestre2022May} (schematically shown in Fig.~\ref{fig: growth}(b)) during a fast cooling after a dwelling time of 2 hours at $\mathrm{1800^{\circ}C}$ and a pressure of 180 MPa under Ar atmosphere. 
The temperature and the pressure are increased at a rate of $\mathrm{100^{\circ}C/min}$ and 10 MPa/min. 
The chamber is initially purged three times (Ar filling followed by pumping) to remove oxygen and moisture.
The sample obtained is an ingot composed of hBN crystals embedded in a solidified $\mathrm{Li_3BN_2}$ matrix. 
$\mathrm{Li_3BN_2}$ dissolution is then performed to retrieve individual crystals. 
They show a lateral size ranging from several hundreds of micrometers to few millimeters, exemplary shown in Fig.~\ref{fig: growth}(b), in the right image.
The crystals have been previously used  as encapsulants for TMDs and graphene to obtain optical and electronic devices \cite{Maestre2022May, Schmitt2023Apr}.

\subsection*{Chemical Vapor Deposition (CVD)}
For the growth of hBN layers, Pt(111) thin films with a thickness of $\mathrm{500 \,nm}$ were prepared on sapphire wafers~\cite{Verguts2017Oct}. 
The hBN films were grown via CVD in an ultra-high vacuum cold-wall chamber for wafers up to 4-inch~\cite{Hemmi2014Mar,Hemmi2021Aug}. 
Prior to all hBN preparations, the Pt/sapphire substrates were cleaned by a series of argon sputtering, $\mathrm{O_2}$ exposure and annealing cycles to $\mathrm{1200\, K}$ until sharp Pt(111) (1$\times$1) LEED patterns were observed. 
Subsequently, hBN layers were prepared at temperatures above $\mathrm{1000 \, K}$ with borazine $\mathrm{(HBNH)_3}$ as precursor with a partial pressure of $\mathrm{10^{-7} mbar}$ (Fig.~\ref{fig: growth}(c)). 
The quality of grown hBN layers were evaluated with scanning low energy electron diffraction (x-y LEED), x-ray photoelectron spectroscopy (XPS), ultraviolet photoelectron spectroscopy (UPS), scanning tunneling microscopy (STM) and atomic force microscopy (AFM). The reported thickness is derived from XPS intensity values.

The transfer procedure employs the electrochemical “bubbling” method~\cite{Cun2018Feb}.
First, the hBN/Pt(111) sample was spin-coated with $\mathrm{4  \,wt \%}$ polymethyl methacrylate (PMMA) ($\mathrm{495 \, K}$).
Then we put the PMMA/hBN/Pt sample as working electrode and a Pt wire as counter electrode in a $\mathrm{1.0\,M}$ KCl solution. 
A negative voltage between $\mathrm{-3\, V}$ and $\mathrm{-5 \, V}$ was applied to the sample to delaminate the hBN/PMMA film from the substrate. 
The delaminated hBN/PMMA film was then rinsed in ultrapure water (Milli-Q Advantage A10) and transferred onto a clean $\mathrm{280 \, nm}$ Si/$\mathrm{SiO_2}$ substrate with gold markers. 
In the next step, the PMMA was removed via a sequence of acetone/ethanol baths and gradual annealing in air at temperatures up to $\mathrm{600 \, K}$ for $\mathrm{3\, h}$ (see transferred BN film in Fig.~\ref{fig: growth}(c)).

\subsection*{Physical Vapor Deposition (PVD)}
Thin nanocrystalline BN films are grown via physical vapor deposition using an ion beam assisted deposition process (IBAD-PVD).
The films exhibit hexagonal bonding structure, as assessed by X-ray absorption near edge spectroscopy (XANES)~\cite{Jimenez1997May, Caretti2011Jul} but lack of x-ray diffraction.
The films were grown directly on $\mathrm{Si/SiO_2}$ wafers with an oxide thickness of $\mathrm{285 \, nm}$ and pre-defined Cr/Au marker.
The growth was performed at room temperature using nitrogen gas and a solid boron source.
The IBAD process consisted in the interaction of a directional beam of $\mathrm{500 \, eV}$ nitrogen ions from a Kaufman source, with concurrent boron atoms from an electron beam evaporator.
The growth setup is schematically depicted in Fig.~\ref{fig: growth}(d).
The N-ion and B-atom fluxes were carefully tuned to obtain stoichiometric BN and avoid other B$_x$N$_y$ phased and bonding configurations~\cite{Caretti2011Aug}.
The thickness of the resulting BN film is $\mathrm{30 \, nm}$ and homogenously covers the whole wafer, as shown in Fig.~\ref{fig: growth}(d).

\section{Cathodoluminesence and Raman spectroscopy on BN crystals and films}

The as-grown hBN crystals (APHT and PCF) are examined by means of TRCL measurements and spatially-resolved confocal Raman microscopy. 
Raman spectroscopy offers a fast and non-invasive tool to spatially probe optical phonons and their lifetimes, which provide a measure of the crystallinity of hBN \cite{Schue2016Dec}.
Time-resolved cathodoluminescence (TRCL) measurements allow to determine the lifetimes of the free excitons, which are strongly affected by scattering with defects, providing a valuable and sensitive tool to locally probe the crystal quality of hBN crystals.
\subsection*{Cathodoluminescence}
\begin{figure}[t]
  \begin{center}
  \includegraphics{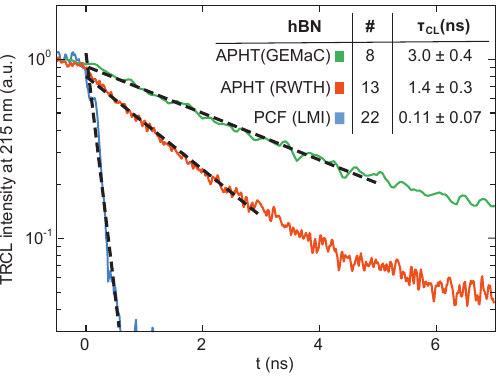}
  \end{center}
  \caption{Decay of the free exciton luminescence at $\mathrm{215\, nm}$ for  hBN crystals grown by APHT (RWTH and GEMaC) and PCF (LMI)  measured by time-resolved cathodoluminescence. 
  The luminescence intensity is normalized to 1  at $t=0$ for all measurements. 
  A representative trace is shown for each crystal type together with an exponential fit to extract the respective lifetime $\mathrm{\tau_{CL}}$.
  The table in the inset shows the averaged lifetimes from measurements at different crystal positions and batches with the total number of measurements.}
  \label{fig: crystals_CL}
\end{figure}

\begin{figure*}[t]
  \begin{center}
  \includegraphics{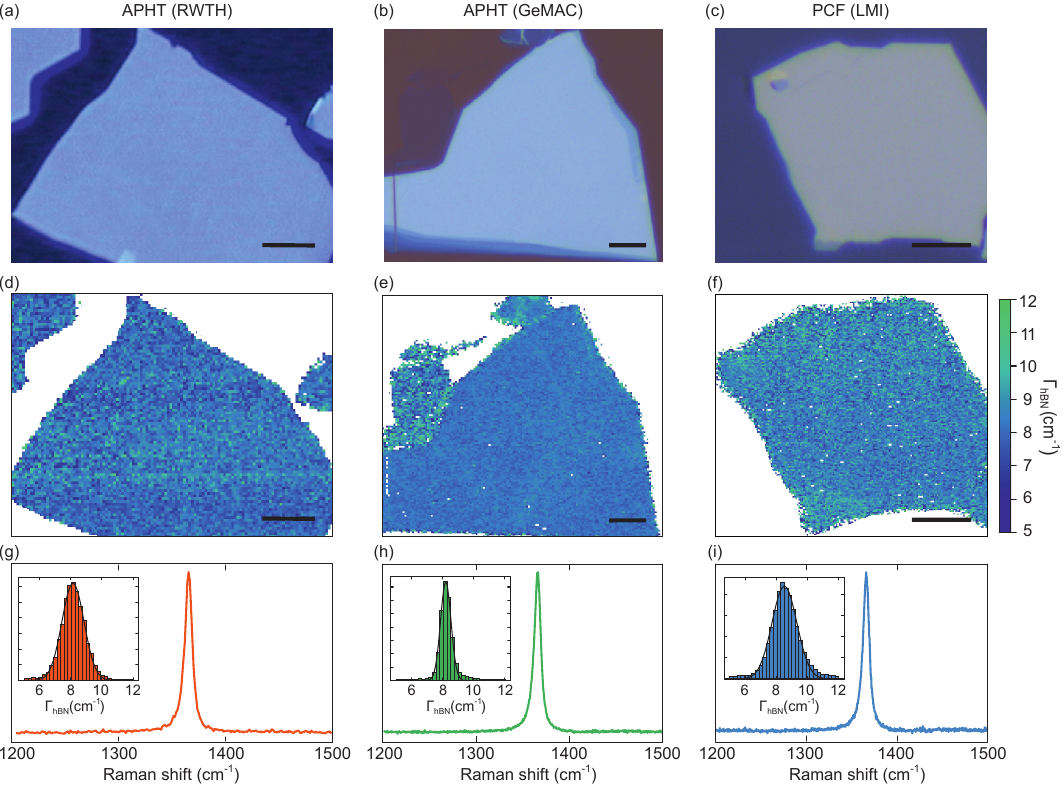}
  \end{center}
  \caption{Raman spectroscopy of hBN crystals for APHT (RWTH and GEMaC) and PCF (LMI) from left to right. {\bf(a) - (c)} Optical images of exfoliated hBN flakes that were spatially mapped by confocal Raman spectroscopy. {\bf(d) - (f)} Spatially-resolved Raman maps of the FWHM of the $E\mathrm{_{2g}}$ mode for each flake shown in (a) - (c). {\bf(g) - (i)} Representative Raman spectra with the statistical distribution of the FWHM of each flake ((d) - (f)) as shown in the respective insets. The scale bars are $\mathrm{10 \, \mu m}$.}
  \label{fig: crystals}
\end{figure*}

TRCL measurements were performed on isolated bulk hBN crystals. For each supplier (GEMaC, RWTH, LMI), crystals from multiple growth batches were investigated. The deep UV spectra were recorded at room temperature in a JEOL7001F field-emission-gun scanning electron microscope (SEM) coupled to a Horiba Jobin-Yvon cathodoluminescence (CL) detection system, as described in detail in earlier works \cite{Schue2016Mar, Schue2019Feb, Roux2021Oct}. 
To allow for time resolution, a custom-built fast-beam blanker was installed inside the SEM column, as described in~\cite{Roux2021Oct}. 
The dynamics of the free exciton population is captured by measuring the time-dependent CL intensity in a wavelength range of $\mathrm{215 \pm 7.5 \, nm}$ with a temporal resolution of $\mathrm{100 \, ps}$. 
This spectral range corresponds to the main luminescence feature of high quality hBN crystals.
The $\mathrm{215 \, nm}$ CL signal results from the indirect exciton recombination assisted via optical phonons. 
To focus on bulk properties and minimize surface recombinations, the electron beam acceleration voltage was set to $\mathrm{15\, kV}$~\cite{Schue2016Mar, Roux2024Apr}.
The current was maintained at a low value of $\mathrm{85\, pA}$ to prevent nonlinear effects~\cite{Plaud2019Jun}. 
An exemplary TRCL measurement for each type of hBN crystal investigated is shown in Fig.~\ref{fig: crystals_CL}. 
At $t=0$, the luminescence peak intensity is normalized to 1, to allow for better  comparison of the time evolution. 
The free exciton lifetime $\mathrm{\tau_{CL}}$ is extracted by fitting the initial decay curve with an exponential function.
We obtain $\mathrm{\tau = 0.15 \, ns, 1.67 \, ns }$ and $\mathrm{3.32 \, ns }$ for PCF (LMI), APHT (RWTH) and APHT (GEMaC) crystals, respectively.
Statistical evaluation (mean and standard deviation) across different growth batches and spatial positions on the crystals yielded $\mathrm{\tau = (0.11 \pm 0.07) \, ns, (1.4 \pm 0.3) \, ns \, and \, (3.0 \pm 0.4) \, ns }$ for 22, 13 and 8 measured areas on PCF (LMI), APHT (RWTH) and APHT (GEMaC), respectively. 
We emphasize the need for statistical evaluation due to notable crystal-to-crystal variations. 

The variation in free exciton lifetimes is associated with differences in defect densities in the crystals.
We note that APHT-grown crystals exhibit lifetimes similar to those produced via the HPHT method \cite{Roux2021Oct} which is consistent with a low defect density. In contrast, PCF-grown crystals show significantly shorter lifetimes. The higher defect density could  
result from vacancies, impurities, or structural anomalies, which all may affect the free lifetime.
The variations in lifetimes, even within crystals grown by the same method, highlights the need for careful crystal selection for specific experiments or applications.
Understanding these defect-induced changes in the optical properties is crucial for the further development of hBN applications in optoelectronics and quantum technology.
The benchmarking of hBN crystals via TRCL also sets the stage for understanding their role as substrates in graphene-based devices.
The observed variation in the exciton lifetime of hBN grown by the different methods is expected to correlate with the electronic quality of encapsulated 2D materials. 
Its impact on the charge carrier mobility in hBN/graphene/hBN Hall bar devices will be detailed further below.

\subsection*{Confocal Raman spectroscopy}
Raman spectroscopy is a practical and widely used optical probe for characterizing both hBN crystals and thin films.
Its advantage of accessibility makes it an important tool for monitoring the effect of changes in the growth parameters on the crystal quality of BN.
The primary benchmark for assessing the crystal quality of hBN via Raman spectroscopy is the FWHM $\Gamma\mathrm{_{E_{2g}}}$ of the $E\mathrm{_{2g}}$ Raman peak, which correlates with the lifetime $\tau\mathrm{_{E_{2g}}}$ of optical phonons corresponding to intralayer vibrations of B and N atoms~\cite{Geick1966Jun}.
The contributions to the phonon linewidth in hBN with a natural isotopic content of boron originate primarily from isotopic disorder-induced scattering, anharmonic phonon decay, or impurity scattering~\cite{Cusco2020Apr}.
Thus, in hBN with the same crystal structure and isotope distribution, the variations in FWHM are mainly due to the degree of disorder in the crystal~\cite{Schue2016Dec}.
Changes in bond lengths due to increased defect density or not-purely $\mathrm{sp^2}$-hybridized bonds might also impact the FWHM, as these factors contribute to averaging effects over phonons of different frequencies.
Typically, high quality hBN crystals grown via HPHT or APHT exhibit a FWHM around $\mathrm{8 \, cm^{-1}}$ \cite{Schue2016Dec, Stenger2017Jun, Gorbachev2011Feb}.
For thin BN films this value can increase up to $\mathrm{40 \, cm^{-1}}$ \cite{Schue2016Dec}. 

\subsubsection*{Experimental setup}
Raman measurements were conducted using a commercial confocal micro-Raman setup (WITec alpha 300R) at room temperature.
We utilized a $\mathrm{532\, nm}$ excitation wavelength, a laser power of  $\mathrm{2\, mW}$ and a 100$\times$ magnification objective with a numerical aperture of 0.9.
The inelastically scattered light was collected through a fibre (core diameter $\mathrm{100 \, \mu m}$) and sent to a CCD through a half meter spectrometer equipped with a $\mathrm{1200 ~\, lines/mm}$ grating.
For linewidth analysis of high-quality hBN crystals, we employed a grating with $\mathrm{2400~\, lines/mm}$.
\subsubsection*{hBN crystals}
We start by exfoliating thin hBN flakes from the bulk crystals using tape (Ultron 1007R) onto  $\mathrm{Si/SiO_2}$ wafer with a $\mathrm{90~\, nm}$ oxide layer and observe similar distribution of thicknesses and lateral sizes for flakes from all crystal suppliers. 
Flakes with thicknesses between $\mathrm{20\, nm}$ and $\mathrm{40\, nm}$ were selected based on their color contrast towards the substrate~\cite{Uslu2024Feb}, as these thicknesses are optimal for building state-of-the-art hBN-encapsulated graphene devices. 
In Figs.~\ref{fig: crystals}(a)-(c), we present optical images of representative flakes from the three suppliers ((a) for APHT (RWTH), (b) for APHT (GEMaC), and (c) for PCF (LMI)). 
All flakes look similar in terms of contamination or thickness homogeneity.
The FWHM of the $E\mathrm{_{2g}}$ peak is extracted by fitting a single Lorentzian function to the individual Raman spectra.
Spatially-resolved maps of the FWHM are shown in Fig.~\ref{fig: crystals}(d)-(f).
A respective single Raman spectrum at a representative position, along with the corresponding histogram to the FWHM map, are shown in Figs.~\ref{fig: crystals}~(g)-(i). 

There is a narrow hBN Raman peak around $\mathrm{1365 \, cm^{-1}}$ for all flakes.
The maps in Figs.~\ref{fig: crystals} (d)-(f) reveal a homogenous and narrow distribution of the FWHM, suggesting uniform crystal quality throughout the exfoliated flakes.
A closer inspection of the statistical distribution (insets of (g)-(i)), reveals a Gaussian distribution of the FWHM around $\mathrm{8 \, cm^{-1}}$, demonstrating high crystallinity for all flakes.
These values are comparable to previous studies on APHT or HPHT grown hBN~\cite{Schue2016Dec}.
Interestingly, we observe no significant difference in the Raman FWHM between PCF and APHT crystals. This observation seems surprising since the CL lifetime of the PCF-grown hBN flakes is more than an order of magnitude shorter than the respective lifetimes of the APHT-grown hBN crystals (see Fig.~\ref{fig: crystals_CL}). It is, however, important to emphasize again that main contributions to the $E\mathrm{_{2g}}$ peak's FWHM in natural hBN results from isotopic disorder~\cite{Cusco2020Apr}, that is typically the same for all. While isotopic disorder is generally the same for all hBN crystals, variations in defect type and density can significantly vary between different growth methods. Our studies suggest that the presence of crystal defects in high-quality hBN crystals can barely be probed by Raman spectroscopy. Analyzing the lifetimes of free excitons, on the other hand, offers a significantly more sensitive tool for the local probing of crystal defects.

\begin{figure*}[t]
  \begin{center}
  \includegraphics{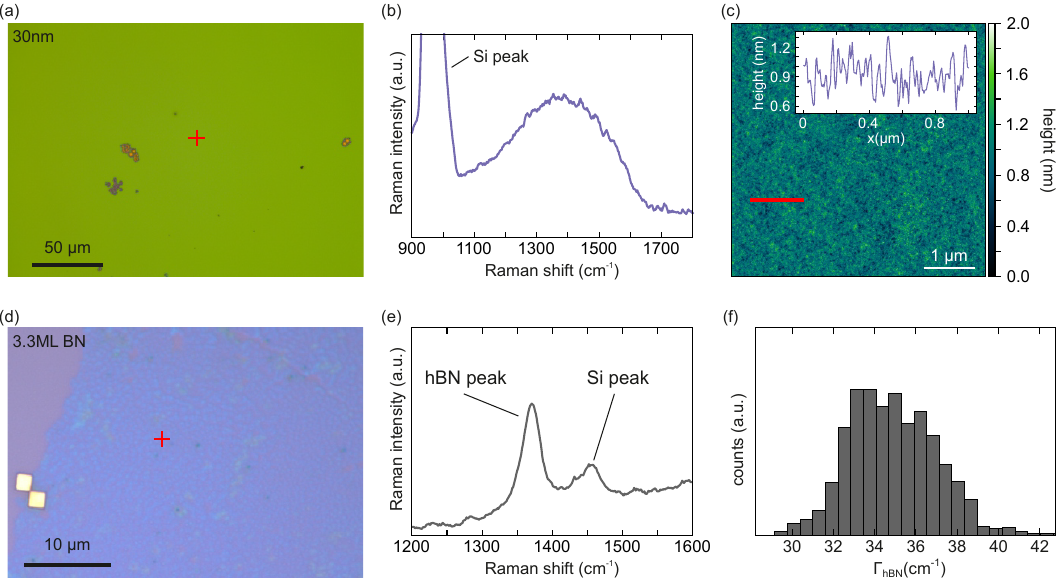}
  \end{center}
  \caption{Boron nitride films on $\mathrm{Si/SiO_2}$. 
  {\bf(a)} Optical microscope image of a PVD-grown BN film with a thickness of $\mathrm{30 \, nm}$. 
  {\bf(b)} Representative Raman spectrum of the PVD-grown BN film (position: red cross in (a))
  {\bf(c)} Surface topography measured by atomic force microscopy. Inset shows a representative line profile along the red horizontal line in the map.
  {\bf(d)} Optical microscope image of a CVD-grown BN film which was transferred to $\mathrm{Si/SiO_2}$ 
  {\bf(e)} Representative Raman spectrum of the CVD-grown BN film (position: red cross in (d))
  {\bf(f)} Histogram of the FWHM of the $\mathrm{E_{2g}}$ hBN mode shown in (d) extracted from single Lorentzian fit.}
  \label{fig: BN_films}
\end{figure*} 

\begin{figure}[t]
  \begin{center}
  \includegraphics{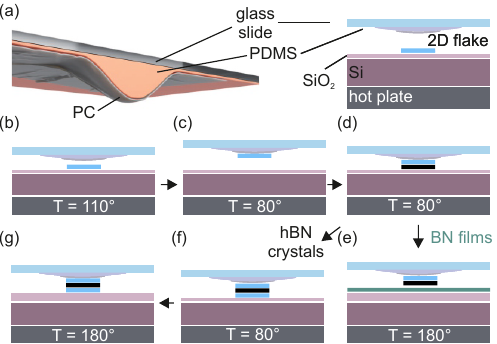}
  \end{center}
  \caption{Dry transfer of graphene/BN heterostructures. 
  \textbf{(a)} Schematic representation of the used stamp. 
  A PC film is placed on a self-assembled PDMS droplet on a glass slide. The stamp is placed above the silicon wafer which is placed on a heatable stage. 
  \textbf{(b)-(c)} The process starts with the pick-up of a hBN flake at $T =\mathrm{ 80 - 110 \, ^{\circ}C}$. 
  \textbf{(d)} The hBN flake is used to pick-up the exfoliated graphene flake at $T =\mathrm{ 80 \, ^{\circ}C}$. 
  Depending on the type of boron nitride, the BN/graphene is then either placed on the BN film on the final substrate \textbf{(e)} or an additional hBN crystal flake is first picked up \textbf{(f)} and then placed on the final substrate \textbf{(g)}. 
  In both cases (e) and (g) the heterostructure is released on the substrate by heating the stage to $T =\mathrm{ 180 \, ^{\circ}C}$ in order to detach the PC from the PDMS and "glue" it to the substrate.
  }
  \label{fig: stacking}
\end{figure}

\begin{figure}[t]
  \begin{center}
  \includegraphics{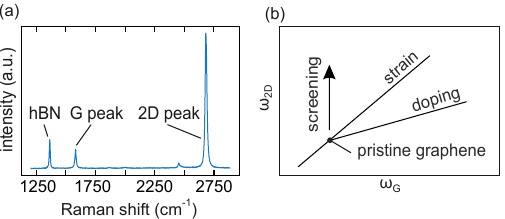}
  \end{center}
  \caption{Raman spectrum of graphene and influence on peak positions. 
  \textbf{(a)} Raman spectrum of graphene encapsulated in hBN. 
  \textbf{(b)} Schematic presentation of the expected influence of strain, doping and screening on the positions of the Raman G and 2D peak of graphene. 
  }
  \label{fig: Raman_basics}
\end{figure}

\begin{figure*}[hbt!]
  \begin{center}
  \includegraphics{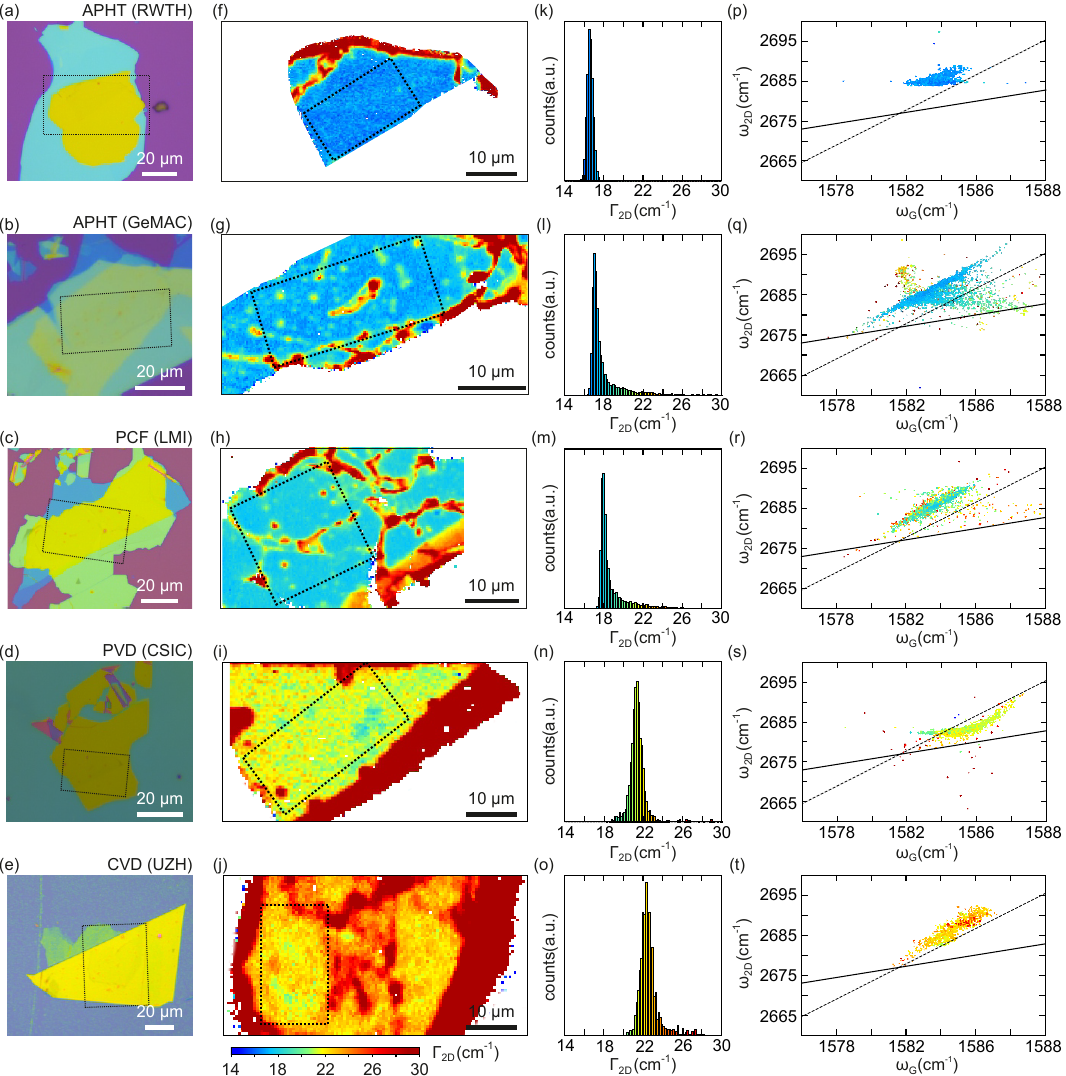}
  \end{center}
  \caption{Optical microsopce images and Raman spectroscopy of dry-transferred hBN/graphene heterostructures. 
  {\bf(a)-(e)} Optical microscope images of one representative stack for each BN source, namely the APHT-hBN from RWTH and GEMaC, the PCF-hBN from LMI, the PVD grown BN film from CSIC and the transferred CVD hBN from UZH. 
  {\bf(f)-(j)} Spatially-resolved Raman map of the 2D FWHM of graphene of the region highlighted as a dashed black rectangle in the optical images in (a)-(e). 
  {\bf(k)-(o)} Statistical representation of the 2D FWHM extracted from the region highlighted as a dashed rectangle in the corresponding Raman maps shown in (f)-(j).
  {\bf(p)-(t)} The 2D peak position vs G peak position. 
  Each point is color coded with the FWHM of the 2D peak.
  The dashed line corresponds to the expected random strain distribution with a slope of 2.2 (see text), while the solid line corresponds to the expected doping distribution with a slope of 0.7.
  }
  \label{fig: Raman_stacks}
\end{figure*}

\subsubsection*{BN films}
We next evaluate boron nitride films, which are either grown directly on the $\mathrm{Si/SiO_2}$ substrate (PVD) or grown by means of CVD and then wet-transferred to a $\mathrm{Si/SiO_2}$ substrate.
In the case of boron nitride films, cathodoluminescence measurements are not feasible, mainly due to the small thickness of the films.
An optical image of the PVD-grown film is shown in Fig.~\ref{fig: BN_films}(a).
We observe a homogeneously grown film over the entire wafer with some spots where the BN is damaged.
In Fig.~\ref{fig: BN_films}(b) we additionally show a Raman spectrum at a representative position.
In contrast to the previously shown Raman spectra of flakes from exfoliated hBN crystals, we do not observe a single narrow Raman peak.
Instead, a broad response ranging from $\mathrm{1100 \, cm^{-1}}$ to $\mathrm{1600 \, cm^{-1}}$ is observed.
This can be related to the amorphous nature of the BN film, which leads to a strong broadening of the Raman peak  due to the inclusion of nanocrystalline regions within the BN film \cite{Nemanich1981Jun}.
The broadening may also result from random strain effects \cite{Sattari-Esfahlan2023Feb}. They lead to an averaging of different bond lengths between the atoms resulting in a statistical averaging of the Raman response due to variations in the phonon frequencies.

As the PVD grown BN films will later be used as a substrate for graphene, we next explore their surface roughness by atomic force microscopy (AFM).
Figure~\ref{fig: BN_films}(c) displays an AFM image for a small region of the sample shown in Fig.~\ref{fig: BN_films}(a).
A root mean square (RMS) roughness of $\mathrm{0.2 \, nm}$ is extracted from this map.
This low value is in line with RMS values 
of hBN and the 2D semiconductor WSe$_2$, which have proven to be ideal substrates for graphene \cite{Banszerus2017Feb}.

In Fig.~\ref{fig: BN_films}(d) we show an optical image of the CVD grown BN film which was transferred on $\mathrm{SiO_2}$. 
Due to the wet-transfer process and because multiple layers of hBN are transferred on top of each other, the BN film does not have a homogenous thickness. 
From XPS measurements we estimate an average thickness of 3 layers of hBN. 
The Raman spectrum at a representative position is shown in Fig.~\ref{fig: BN_films}(e) together with a histogram of the distribution of the FWHM in panel (f).
We observe a well-defined hBN Raman peak at $\mathrm{\omega_{E_{2g}} = 1365 \, cm^{-1}}$ with a FHWM of $\mathrm{\Gamma_{E_{2g}} = 34 \, cm^{-1}}$.
The large FWHM is in striking contrast to the previously discussed crystals but comparable to other BN films shown in literature \cite{Song2010Aug, Shi2010Oct, Hadid2022Nov}. 
We attribute the large FWHM to the wet transfer procedure and the remaining PMMA residues on the transferred BN film. 

To conclude the pre-characterization of BN crystals and films, we note that there is no common method which is either sensitive enough or applicable to all forms of BN, i.e. crystals and films.
Especially, for nanocrystalline or amorphous BN films, which are recognized as potential substrates for scaled devices, the usual characterization methods are not feasible.
We therefore proceed with the evaluation of graphene in contact with BN, by using graphene as a sensitive detector for the suitability of the underlying BN/hBN substrate for charge transport.

\section{Dry-transfer of graphene encapsulated in BN}
The next step in the benchmarking protocol is to build van der Waals heterostructures using BN material to fully encapsulate graphene. The substrate quality of BN is then explored by probing the electronic properties of graphene using both spatially-resolved Raman spectroscopy and charge transport measurements. 

For the stacking of the heterostructures we start by exfoliating hBN and graphene flakes onto $\mathrm{90\, nm}$ $\mathrm{Si/SiO_2}$. 
The flakes are searched and classified using a home-built automatic flake detection tool \cite{Uslu2024Feb}. 
Suitable flakes with a thickness between 20 and 40 nm are identified and stacked on top of each other using standard dry-transfer methods with  poly(bisphenol A carbonate) (PC) film on top of a drop-shaped polydimethylsiloxane (PDMS) stamping tool \cite{Bisswanger2022Jun}. 
The stacking process is schematically depicted in Fig.~\ref{fig: stacking}. 
For the benchmarking of hBN crystals (APHT and PCF), graphene is picked up using hBN flakes, which were exfoliated from their respective bulk crystals while for the evaluation of BN films the graphene is picked up by exfoliated HPHT-grown hBN (Figs. \ref{fig: stacking}(b)-(d)).  
In the next step, the hBN/graphene half stack is either transferred onto corresponding hBN crystal flakes (Figs. \ref{fig: stacking}(f)-(g)) or placed onto the BN films (Fig.~\ref{fig: stacking}(e)).
The protection of graphene from the top by an hBN crystal is important to ensure heterostructures of comparable quality and exclude influences on the graphene quality and device performance that can be caused by chemicals or airborne contaminations \cite{Palinkas2022Nov} during the subsequent processing steps.
Optical microscope images of the finished stacks are shown in Fig. \ref{fig: Raman_stacks}(a)-(e). 
The lateral size of the stacks is limited by the size of the exfoliated hBN and graphene flakes. 
Within this project, we characterized in total over 40 dry-transferred samples to obtain a statistical evaluation of the various BN substrate and to exclude sample-to-sample variations. 

\section{Raman spectroscopy on BN-graphene heterostructures}

\subsection*{Extraction of strain, strain variations and doping}
We first give an overview on the key concepts of graphene-based Raman spectroscopy.
Figure \ref{fig: Raman_basics}(a) shows a typical Raman spectrum of graphene encapsulated in hBN crystals. 
Three prominent peaks are typically observed corresponding to the above analyzed hBN $\mathrm{E_{2g}}$ peak and the graphene G and 2D peak. 
The G peak in graphene results from out-of-phase in-plane vibrations of two carbon atoms of the two sublattices and involves phonons from the $\mathrm{\Gamma}$-point, whereas the double resonant 2D peak corresponds to a breathing mode, involving phonons near the K-point \cite{Ferrari2006Oct,Graf2007Feb,Ferrari2013Apr}.

A crucial and sensitive quantity for the evaluation of the electronic properties of graphene is the FWHM of the 2D peak, which is directly connected to the extent of nm-scale strain variations within the laser spot~\cite{Neumann2015Sep} and therefore also contains information on the roughness of the substrate~\cite{Banszerus2017Feb}. 
As strain variations locally break the hexagonal symmetry of the lattice, a vector potential is induced which in turn leads to an increased probability of backscattering of electrons in charge transport leading to a reduced charge carrier mobility~\cite{Couto2014Oct}. 
The 2D FWHM is therefore the main quantity of interest in our study as it directly connects the interface quality given by the BN with the electronic quality of the adjacent graphene sheet.

The G and 2D peak are both susceptible to strain as well as doping \cite{Lee2012Aug} and the 2D peak position is additionally influenced by dielectric screening from the environment \cite{Forster2013Aug}, which is, however, not relevant in the scope of this study. 
To separate the effects of strain and doping from spatially-resolved Raman maps, the positions of the 2D and G peak are plotted against each other, as illustrated in Fig.~\ref{fig: Raman_basics}(b). 
Since the two peaks shift differently as function of doping and strain, the slopes of the distributions can be used to qualitatively evaluate the type of disorder in the system (strain and/or doping). 
A distribution parallel to the strain axis has a slope of 2.2 and is connected to biaxial strain whereas a distribution along the doping axis has a slope ranging between 0.3 and 0.7 depending on both their charge carrier type and the substrate~\cite{Lee2012Aug,Sonntag2023Feb}.

\subsection*{Results of spatially-resolved Raman spectroscopy}
Raman measurements were performed with the same setup as for the characterization of the hBN crystals and films, using a grating of $\mathrm{1200\,lines/mm}$.
\begin{figure}[t]
  \begin{center}
  \includegraphics{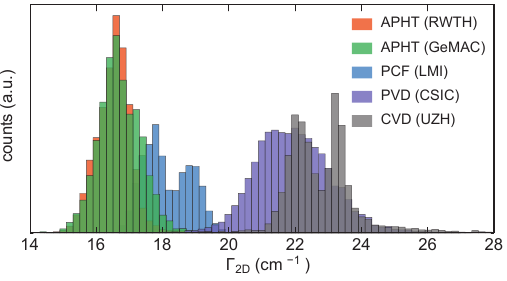}
  \end{center}
  \caption{
  Distributions of the graphene 2D peak FWHM for all fabricated and evaluated heterostacks, combined in a single histogram for each BN source.} 
  \label{fig: Raman_FWHM}
\end{figure}
\begin{figure*}[t]
  \begin{center}
  \includegraphics{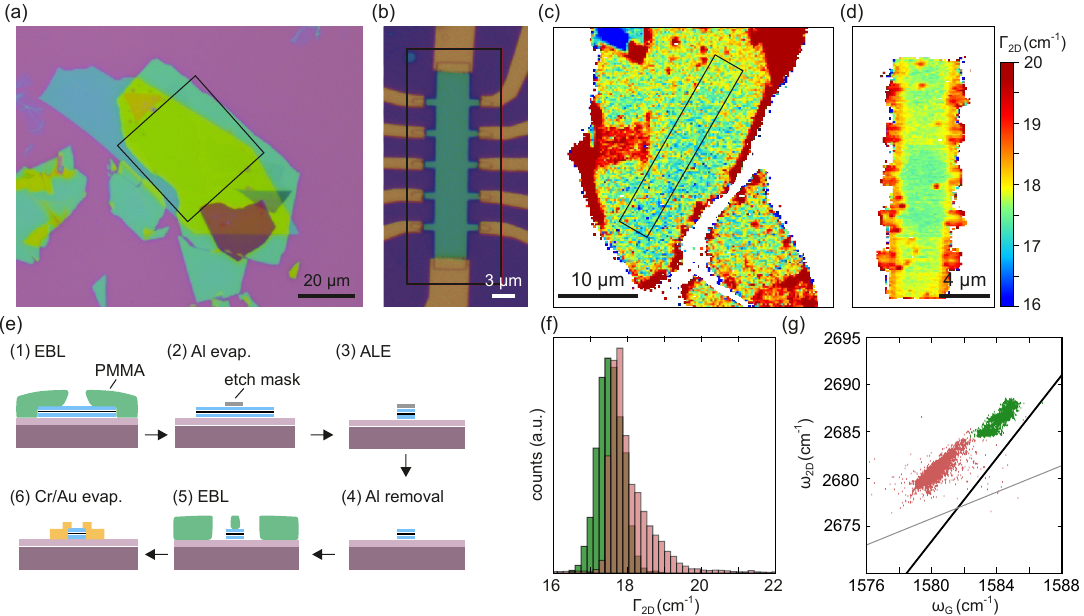}
  \end{center}
  \caption{Fabrication of Hall-bar structures from van der Waals (vdW) heterostructures. 
  {\bf(a)} Optical microscope image of hBN/graphene/hBN vdW heterostructure. The black rectangle denotes the area mapped by  Raman spectroscopy shown in panel {\bf (c)}. 
  {\bf (b)} Optical microscope image of the patterened and contacted Hall bar.   
  {\bf(c)} Spatially-resolved Raman map of the graphene 2D FWHM. The black rectangle corresponds to the position where the Hall bar is placed. 
  {\bf(d)} Spatially-resolved Raman map of the 2D FWHM of the finished Hall bar. 
  {\bf(e)} Schematic of the process overview for Hall bar structures. (1) Electron beam lithography to define the Hall bar structure, followed by (2) electron beam evaporation of aluminum and (3) subsequent atomic layer etching. After chemical etching of the aluminum in TMAH (4) the contacts are defined in a second EBL step (5) and the Hall bar is finally contacted by Cr/Au evaporation (6). 
  {\bf(f)} Histogram of the Raman 2D peak FWHM before (green) and after (red) Hall bar fabrication.
  {\bf(g)} Scatter plot of the graphene 2D vs G peak position before and after Hall bar fabrication. The black line shows the expected distribution for biaxial strain (slope = 2.2) and the grey line for doping (slope = 0.7).
  }
  \label{fig:processing}
\end{figure*}
\begin{figure*}[t]
  \begin{center}
  \includegraphics{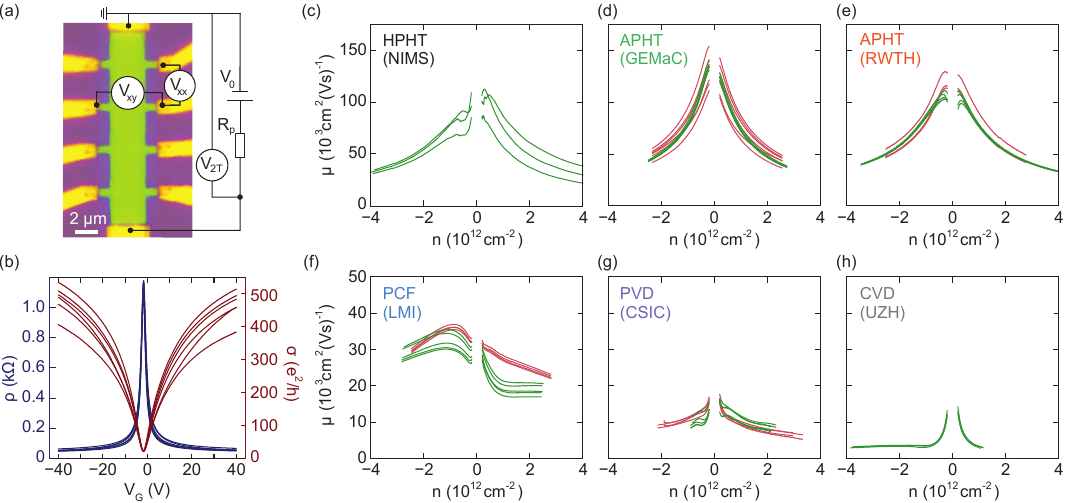}
  \end{center}
  \caption{Charge transport measurements on graphene/BN Hall bars. 
  {\bf (a)} Optical image of a representative Hall bar structure with a schematical representation of the electrical wiring.
  {\bf (b)} Four-terminal resistivity (conductivity) as function of the silicon back gate voltage.
  {\bf (c)-(g)} Extracted Drude mobilities as function of charge carrier density for HPHT-NIMS, APHT-RWTH, RWTH-GEMaC, PCF-LMI, PVD-CSIC and CVD-UZH (wet-transferred), respectively.
  Traces of the same color correspond to multiple regions measured within the same device.  
  Different colors correspond to different devices.
  }
  \label{fig:mobilities}
\end{figure*}
Figures~\ref{fig: Raman_stacks}(f)-(j) show Raman maps of the graphene 2D linewidth for the regions highlighted with black dashed rectangles in the optical images in Figs.~\ref{fig: Raman_stacks}(a)-(e) for each BN source, respectively. 
The corresponding histograms are shown in Figs.~\ref{fig: Raman_stacks}(k)-(o) of a selected region of interest, highlighted with a dashed rectangle in the corresponding panel in Figs.~\ref{fig: Raman_stacks}(f)-(j).
The color scale is the same for all maps. 
Regions of a higher 2D linewidth within a stack may either result from bubbles (hydrocarbons) that are trapped at the interface between hBN and graphene or may be related to regions with multilayers. 
Residual hydrocarbons most likely originate from tape residues during  exfoliation or from the polymer used for  stacking \cite{Lin2012Jan,Schwartz2019Jul,Volmer2021Mar}. The latter is a commonly known challenge when using  polymer-based dry-transfer techniques.
We observe the formation of bubbles for all stacks produced in this study.

Comparison of the contamination-free regions of the 2D FWHM maps reveals the lowest $\mathrm {\Gamma_{2D}} $ values for graphene on APHT-hBN (Figs.~\ref{fig: Raman_stacks}(f)-(g)), followed by PCF-grown crystals (Fig.~\ref{fig: Raman_stacks}(h)) and than the BN films (Figs.~\ref{fig: Raman_stacks}(i)-(j)).
The corresponding histograms in Figs.~\ref{fig: Raman_stacks}(k)-(o) enable the quantitative evaluation of the 2D FWHM maps.
The maximum of the statistical distribution ranges from $\mathrm{16.5 \, cm^{-1}}$ for APHT-grown crystals, over $\mathrm{18 \, cm^{-1}}$ for PCF-grown crystals to values larger than $\mathrm{20 \, cm^{-1}}$ for BN films.
We identify the peak position of the 2D linewidth distribution as a robust and sensitive quantity to evaluate the interface quality of the underlying BN, in line with previous works \cite{Neumann2015Sep, Banszerus2017Feb}.  
We conclude that the degree of strain variations in graphene is lowest for the APHT hBN crystals, which shows that they have the highest interface quality (flatness) among the studied BN. 
The respective $\omega_{\mathrm{2D}}$ vs $\omega_{\mathrm{G}}$ scatter plots are shown in Fig.~\ref{fig: Raman_stacks}(p)-(t), where the color code corresponds to the FWHM of the 2D peak.
For the stack presented in the first row of Fig.~\ref{fig: Raman_stacks} we chose a region with a spatially homogeneous and low 2D FWHM. The corresponding 2D vs G peak position distribution shows a strong clustering along the 2.2 strain axis indicating very small strain variations and negligible doping.
For the sample in the second row, the distribution with the lowest 2D linewidth (blue data points) is again mainly distributed along the strain axis. However, areas with inclusion (bubbles) exhibit larger 2D linewidths (green, yellow and reddish color) with a distribution outside the strain axis, which is probably due to larger doping.
The effect of doping on the peak positions is most clearly seen for the PVD-grown BN shown in the fourth row of Fig.~\ref{fig: Raman_stacks}.
The peak positions show a curved distribution that results from both strain and doping.

To go beyond the evaluation of the comparison of representative examples, we plot the graphene 2D linewidth of high-quality regions of all evaluated samples in a combined histogram in Fig.~\ref{fig: Raman_FWHM}.
For the APHT-grown crystals we observe narrow distributions of the graphene 2D linewidth with the maximum at $\mathrm{16.5 \, cm^{-1}}$, demonstrating  an excellent and reproducible interface quality between graphene and hBN  over a number of 20 different heterostructures with hBN crystals taken from different batches. 
The histogram distribution of the PCF-crystals shows a broader distribution ranging from $\mathrm{17.5 \, cm^{-1}}$ to $\mathrm{19 \, cm^{-1}}$ indicating a larger amount of strain variations, and when evaluating different stacks, we additionally observe a larger sample-to sample variation in the 2D linewidth distribution.

While the analysis of the free exciton lifetime $\mathrm{\tau_{CL}}$ in Fig.~\ref{fig: crystals_CL} shows slightly shorter lifetimes for RWTH-APHT crystals compared to the GEMaC-APHT crystals there are no differences in the amount of nm-strain variations of encapsulated graphene as inferred from Raman spectroscopy. In contrast, the broader and shifted graphene 2D linewidth distribution of heterostacks fabricated by the PCF crystals seems to be related to their shorter exciton lifetimes. 
As the graphene 2D linewidth is connected to nm-strain variations caused by the roughness of the substrate surface, we conclude that the defect concentration in the PCF-grown crystals is so high that it affects the electronic properties of graphene.
Further, this quantity allows us to compare various substrates independent on their crystal nature to each other.

\section{Processing into Hall Bar structures}

We next determine the key quantity of interest, the charge carrier mobility of graphene, and link it to the Raman 2D linewidths of graphene and the free exciton lifetimes of the BN substrate.
For this purpose, the fabricated heterostructures are patterned into Hall bar devices and electrically contacted to perform gate-dependent charge transport measurements.
For this study, we established a reproducible fabrication process yielding a high homogeneity of the electronic quality of graphene within a device as well as a high throughput of functioning contacts. 
For all devices we applied the same fabrication routine. 

A simplified overview of the various processing steps is shown in Fig.~\ref{fig:processing}(e). 
First, the Hall bar structure is defined by electron beam lithography (EBL) (step 1). 
Subsequently, $\mathrm{30 \, nm}$ aluminum (Al) is deposited using electron beam evaporation with a rate of $\mathrm{0.1 \, nm/s}$ (step 2) and after lift-off we remain with the final Hall-bar structure protected by the Al hard mask (step 3).
The structure is subsequently etched using atomic layer etching (Oxford Plasma Pro 100) using Ar/$\mathrm{SF_6}$ with a flow rate of $\mathrm{5/20~sccm}$ and HF power of $\mathrm{50 \, W}$ and a 5s oxygen etch pulse. 
The Al is chemically removed using tetramethylammonium hydroxide (TMAH) (step 4). 
The contacts to the Hall bar are defined in a second EBL step (step 5) and $\mathrm{5\, nm/70\, nm}$ of Cr/Au is evaporated, with a rate of $\mathrm{0.2 \, nm/s}$ and $\mathrm{0.5 \, nm/s}$ (step 6). 
An optical microscope image of a representative, structured and contacted device is shown in Fig.~\ref{fig:processing}(b). 

At this point, it is important to note that we have taken particular care to minimize the time between the individual processing steps. 
The etching, the subsequent second lithography step and the evaporation of Cr/Au was performed within the same day. 
By fabricating many devices, we have clear evidence that the time window between etching into the Hall bar structure where we expose the edges of graphene to air and the deposition of the side contacts to graphene should be minimized. For all devices, this time window was below $\mathrm{4 \, h}$.

\subsection{Influence of processing on the electronic properties of graphene}

In this section we discuss the impact of the Hall bar processing onto the mechanical and electronic quality of the devices by using spatially-resolved Raman spectroscopy.
In Fig.~\ref{fig:processing} we show a representative device PCF (LMI), with an optical image of the stack in panel (a) and the final device in panel (b). 
Figures~\ref{fig:processing}(c) and (d) depict spatially-resolved Raman maps of the graphene 2D linewidth of the heterostructure before and after processing, respectively. 
The black rectangle in Fig.~\ref{fig:processing}(c) illustrates the region chosen for the Hall-bar patterning, and only the Raman data from this region are used for comparison with the final Hall bar.
The respective histogram is shown in Fig.~\ref{fig:processing}(f) (green data).
A comparison of the two maps in Figs.~\ref{fig:processing}(c) and (d) shows: (i) an overall increases in the Raman 2D FWHM in the center of the Hall bar, which leads to a shift of the respective histogram (red data in Fig.~\ref{fig:processing}(f)) towards higher wavenumbers and (ii) a strong increase in linewidth towards the edges of the Hall bar (reddish color in Fig.~\ref{fig:processing}(d) that is seen as a tail in the histogram extending to values above 20 cm$^{-1}$. 
This finding could be linked to mechanical stress that occurs during the fabrication steps.
The different temperatures in the fabrication process, e.g. after baking the resist for lithography or during etching, can lead to stress due to the different thermal expansion coefficients of the materials within the stack and the substrate.

Considering the Raman 2D and G peak positions in Fig.~\ref{fig:processing}(g), we clearly observe a red shift of the positions along the 2.2 strain line for the stack after fabrication.
This cloud (red data points) has shifted towards phonon frequencies closer to the point related to that of "pristine" graphene~\cite{Lee2012Aug}, suggesting that strain release may have occurred  during the device fabrication.
We only show one example here, but this finding is observed in many different samples, regardless of the type of BN used.
A more detailed investigation is beyond the scope of this paper and future works focusing on the monitoring of different fabrication steps are necessary to draw clearer conclusions.

\subsection{Room temperature charge carrier mobilities}
The individual Hall bars with the different BN substrates were fabricated in heterostack regions of the lowest possible and homogeneous graphene Raman 2D FWHM (as an example, see black rectangle  in Fig.~\ref{fig:processing}(c)).
All charge transport measurements were taken at room temperature under vacuum. 
An example of a Hall bar with the measurement scheme is depicted in Fig.~\ref{fig:mobilities}(a). 
We use an AC voltage $V\mathrm{_0 = 1\, V}$ at a frequency of 77~Hz and a series resistance of $R\mathrm{_P=1\,M\Omega}$ to pass a constant current of $I =\mathrm{ 1\, \mu A}$ between the source and drain contact. 
The four-terminal voltage drop is measured for different regions along the graphene transport channel, labelled as $V\mathrm{_{xx}}$ in Fig.~\ref{fig:mobilities}(a) for the upper region as an example .  
This voltage drop converts to the  resistivity (1/conductivity) following $\mathrm{\rho = 1 /\sigma} = W/L \cdot V_\mathrm{{xx}}/I$, where $L$ is the distance between the contacts and $W$ the width of the transport channel. 

Figure~\ref{fig:mobilities}(b) shows the gate dependent resistivity and conductivity for an APHT device (red traces in panel (d)). 
For all measured regions, the conductivity $\mathrm{\sigma}$ reaches at least $\mathrm{400 \, e^2/h}$ at large gate voltages, i.e. large charge carrier densities, which is mainly limited by electron-phonon scattering \cite{Wang2013Nov}. 
Importantly, and in contrast to previous studies, we observe homogeneous transport properties along the graphene channel and a high yield of functioning contacts (larger than 90 $\%$).
While the electronic homogeneity is likely due to the pre-selection of the regions via Raman mapping, we link the high throughput of functioning contacts to the decreased time between the etching (i.e. exposing of graphene contact areas) and evaporation of the Cr/Au.

\begin{figure*}[t]
  \begin{center}
  \includegraphics{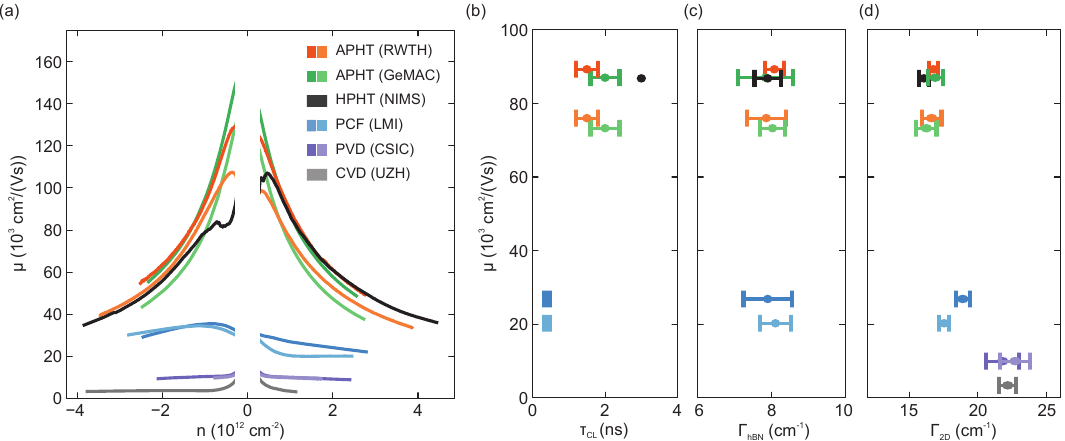}
  \end{center}
  \caption{Summary of the main findings of the BN benchmarking study.
  {\bf (a)} Charge carrier mobility as a function of Drude mobility for the best transport region of each device for every supplier as a function of charge carrier density.
  The charge carrier mobility of a reference device fully encapsulated in HPHT-hBN is also shown for reference.
  Charge carrier mobility at a charge carrier density of $\mathrm{\num{1e12}\, cm^{-2}}$ as function of {\bf (b)} the free exciton lifetime, {\bf (c)} the averaged FWHM of the $E\mathrm{_{2g}}$  Raman peak, and {\bf (d)} the averaged graphene Raman 2D linewidth of the Hall bar after fabrication.
  }
  \label{fig:benchmarking}
\end{figure*}

The charge carrier density $n$ is extracted from Hall effect measurements. It is connected to the gate voltage by $n=\alpha (V_\mathrm{G}-V\mathrm{_G^0})$, where $V\mathrm{_G^0}$ is the position of the charge neutrality point, i.e. the voltage of the Dirac peak, and $\alpha$ is the gate lever arm.  
The carrier mobilities $\mu = \sigma/(ne)$ of graphene with the different BN substrates are shown in Figs.~\ref{fig:mobilities}(c)-(h) for each BN source individually. As a reference, we show transport data for a Hall bar device where we used HPHT hBN (NIMS) (see Fig.~\ref{fig:mobilities}(c)).  
For each device, multiple regions were measured. Traces of the same color are from different regions of the same device. There are only small variations in transport characteristics within a single device but also between different devices fabricated from the same BN source.  
This finding further confirms a robust and reliable processing routine, which was developed as part of the benchmarking study. 
For devices built by APHT hBN we measure the highest charge carrier mobilities exceeding $\mathrm{80,000\, cm^2(Vs)^{-1}}$ at a charge carrier density of $|n|=\mathrm{\num{1e12} \, cm^{-2}}$ (see Figs.~\ref{fig:mobilities}(c) and (d)).  
These values are fully in line with state of the art high-mobility graphene devices using HPHT hBN \cite{Wang2013Nov} (see also Fig.~\ref{fig:mobilities}(c)) or APHT hBN from other sources \cite{Sonntag2020Jun,Onodera2020Jan,Ouaj2023Sep}. 
We therefore highlight the viability of APHT hBN crystals as a true alternative to HPHT hBN crystals, for high-performance graphene devices.
For the PCF-grown crystals in Fig.~\ref{fig:mobilities}(f), we observe a  carrier mobility of up to $\mathrm{30,000\, cm^2(Vs)^{-1}}$ at $|n|=\mathrm{\num{1e12} \, cm^{-2}}$. 
The lower charge carrier mobility of graphene encapsulated in PCF-grown hBN, when compared to APHT hBN, is fully consistent with our two previous observations, a shorter free exciton lifetime and a higher 2D linewidth of graphene encapsulated in PCF-grown hBN crystals.
The relation between an increase in graphene 2D linewidth and a decrease in charge carrier mobility is understood in terms of increased electron backscattering  due to the stronger nm-strain variations~\cite{Couto2014Oct}. 
For the PVD grown BN film (Fig.~\ref{fig:mobilities}(g)) we extract charge carrier mobilities over $\mathrm{10,000\, cm^2(Vs)^{-1}}$ at  $n=\mathrm{\num{1e12} \, cm^{-2}}$, while we achieve mobilities around $\mathrm{4,000\, cm^2(Vs)^{-1}}$ at  $n=\mathrm{\num{1e12} \, cm^{-2}}$ for the CVD-grown and wet-transferred films (Fig.~\ref{fig:mobilities}(h)).

\section{Discussion}

In Fig. \ref{fig:benchmarking}, we summarize the main results of the BN benchmarking study: (a) room temperature charge carrier mobility vs carrier density and carrier mobilities at  $n=\mathrm{\num{1e12} \, cm^{-2}}$ vs (b) free exciton lifetime of hBN, (c) $\Gamma_\mathrm{{E_{2g}}}$ of hBN and (d) graphene 2D linewidth of all BN substrates. 
In Fig. \ref{fig:benchmarking}(a) we show the transport traces for the region of highest mobility for each device shown in Figs.~\ref{fig:mobilities}(c) to (h). 
As mentioned above, APHT grown hBN allows for equally high graphene mobilities as achieved for HPHT-grown hBN crystals. 
These hBN sources are of high relevance for many research groups, who are interested in high quality hBN crystals for fundamental research. 
The PCF-grown crystals, following another route of hBN crystal growth, allow for mobilities up to $\mathrm{30,000\, cm^2(Vs)^{-1}}$ at  $n=\mathrm{\num{1e12} \, cm^{-2}}$, demonstrating the great potential of new synthesis routes for the production of high quality hBN crystals.  
One aim of this synthesis route is to satisfy the increasing demand of hBN crystals from many research groups. 
However, these approaches to grow high quality hBN crystals are not scalable, because they cannot be easily combined with technologically relevant substrates and the desired thicknesses can only be achieved via mechanical exfoliation. 
Scalable methods for growing BN are therefore needed to unlock the full potential of graphene-based electronics in future nanoelectronic devices. 

In this respect, the PVD growth method is most promising because (i) it allows the growth of tens of nanometer thick films with very low surface roughness and (ii) it can be deposited directly onto $\mathrm{Si/SiO_2}$ substrates. The large BN thickness screens disorder from the silicon substrates, while the deposition on the target substrates prevents the need for large scale layer transfer. Most importantly, the PVD-grown BN allows for room temperature carrier mobilities of graphene exceeding $\mathrm{10,000\, cm^2(Vs)^{-1}}$ at  $n=\mathrm{\num{1e12} \, cm^{-2}}$. We conclude that the low temperature PVD growth process of BN on $\mathrm{SiO_2}$ is a promising platform for achieving scalable BN substrates not only for graphene, but also for other 2D materials. 

If we compare the charge carrier mobility with the Raman 2D FWHM we see a clear trend of decreasing mobility with increasing 2D FWHM. 
This is in good agreement with the finding that nm-scale strain variations are the limitation for high charge carrier mobilities
\cite{Couto2014Oct} and shows that the graphene Raman 2D FWHM is a good measure for benchmarking as used in the ICE TS 62607-6-6 key control characteristics \cite{ICE}. 
Whereas we do not find a correlation between the FWHM of the hBN Raman peak and the charge carrier mobility we observe a clear correlation between the CL lifetime and the mobility, as shown in Fig.~\ref{fig:benchmarking}(b). 
We observe that we need CL lifetimes of over $\mathrm{1 \, ns}$ to achieve charge carrier mobilities in the range of $\mathrm{80,000\, cm^2(Vs)^{-1}}$ at  $n=\mathrm{\num{1e12} \, cm^{-2}}$. 
For CL lifetimes of $\mathrm{100 \, ps}$ we achieve charge carrier mobilities up to $\mathrm{30,000\, cm^2(Vs)^{-1}}$. 
The interface quality is therefore connected to the hBN crystal quality, i.e. the number of defects, in a sensitive way.

In conclusion, we have presented a comprehensive study of the electronic properties of graphene on different boron nitride substrates using a newly developed reproducible processing routine.
We have shown the complete process from boron nitride synthesis, over its  optical characterization, to the optical and electronic characterization of graphene after encapsulation and Hall bar fabrication. 
We identify the Raman spectrum of BN as a valuable measure for distinguishing hBN in the high crystallinity limit from BN films, but we also point out the limitations of the Raman analysis when comparing high-quality hBN crystals. 
In this respect, time-resolved cathodoluminescence has a clear advantage over Raman spectroscopy when evaluating the as-grown quality of hBN, as the probing of the free exciton lifetime is very sensitive to the defects in hBN. 
The fabrication of graphene-based heterostructures on BN  substrates demonstrates the high sensitivity of graphene to the environment, allowing graphene to be used as a sensitive detector of the substrate and interface quality.
Variations in the quality of the graphene-BN interface are directly reflected in a broadening of the graphene Raman 2D peak. This broadening has a direct effect on the carrier mobility, i.e. the mobility is inversely proportional to the peak of the 2D linewidth distribution of graphene. It is therefore advisable to characterize the Raman 2D linewidth distribution of the finished heterostructure prior to any processing.      
In terms of benchmarking we find that a CL lifetime larger than $\mathrm{1\,ns}$ is sufficient for high hBN crystal quality and high graphene-hBN interface qualities with low nm strain variations in graphene, which is essential for fundamental studies on highest mobility graphene-based devices. 
For scalable approaches we see that a graphene Raman 2D linewidth  below $\mathrm{22\, cm^{-1}}$ is necessary to achieve charge carrier mobilities over $\mathrm{10,000\, cm^2(Vs)^{-1}}$. 
PVD-grown BN films, therefore, offer a promising platform for scalable high mobility graphene devices. 
\section{Data availability}
The data supporting the findings of this study are available in a Zenodo repository under, https://doi.org/10.5281/zenodo.13684712.
\section{Acknowledgements}
This project has received funding from the European Union’s Horizon 2020 research and innovation programme under grant agreement No. 881603 (Graphene Flagship), T.O., S.B., P.S, C.S., and B.B. acknowledge support from the European Research Council (ERC) under grant agreement No. 820254, and the Deutsche Forschungsgemeinschaft (DFG, German Research Foundation) under Germany’s Excellence Strategy - Cluster of Excellence Matter and Light for Quantum Computing (ML4Q) EXC 2004/1 - 390534769. H.Y.C. was supported by a SPARK grant of the Swiss National Science Foundation (Grant No. CRSK-2\_220582). A.H. acknowledges a Forschungskredit of the University of Zürich (Grant No. FK-20-206 114). 
K.W. and T.T. acknowledge support from the JSPS KAKENHI (Grant Numbers 21H05233 and 23H02052) and World Premier International Research Center Initiative (WPI), MEXT, Japan.


%


\end{document}